\documentclass[conference]{IEEEtran}
\usepackage{color, soul}
\usepackage{float}

\usepackage{cite}
\usepackage{amsmath,amssymb,amsfonts}
\usepackage{algorithmic}
\usepackage{graphicx}
\usepackage{textcomp}
\usepackage{xcolor}
\usepackage{braket}
\usepackage{quantikz}
\usepackage{balance}
\usepackage[hidelinks]{hyperref}
\usepackage[a4paper, total={184mm,239mm}]{geometry}

\newcommand\copyrighttext{%
  \footnotesize \textcopyright 2021 IEEE.  Personal use of this material is permitted.  Permission from IEEE must be obtained for all other uses, in any current or future media, including reprinting/republishing this material for advertising or promotional purposes, creating new collective works, for resale or redistribution to servers or lists, or reuse of any copyrighted component of this work in other works. DOI: \href{https://doi.org/10.1109/VTS50974.2021.9441047}{10.1109/VTS50974.2021.9441047}}

\usepackage{bbm}
\begin{document}

\title{Special Session: Noisy Intermediate-Scale\\ Quantum (NISQ) Computers---How They Work,\\ How They Fail, How to Test Them?
}

\author{
\thanks{\copyrighttext .}
\begin{tabular}{@{}ccc@{}}
\multicolumn{3}{c}{Sebastian Brandhofer$^1$ \qquad Simon Devitt$^2$ \qquad Thomas Wellens$^3$ \qquad Ilia Polian$^1$}\\[0.5cm]
    \begin{tabular}{c}
\normalsize
    $^1$University of Stuttgart and Center for\\
\normalsize
    Integrated Quantum Science and Technology\\
\normalsize
    Stuttgart, Germany \\
\normalsize
    \{sebastian.brandhofer\;$|$\;ilia.polian\}\\
\normalsize
    @informatik.uni-stuttgart.de
    \end{tabular}
&
    \begin{tabular}{c}
\normalsize
\normalsize
    $^2$University of Technology Sydney\\
\normalsize
    Sydney, Australia \\
\normalsize
    simon.devitt@uts.edu.au\\\hspace{0cm}\\\\
    \end{tabular}
&
    \begin{tabular}{c}
\normalsize
\normalsize
    $^3$Fraunhofer Institute for\\ 
\normalsize
    Applied Solid State Physics IAF \\
\normalsize
    Freiburg, Germany \\
\normalsize
    thomas.wellens@iaf.fraunhofer.de\\\\
    \end{tabular}
\end{tabular}
}
\IEEEoverridecommandlockouts

\IEEEpubidadjcol
\maketitle

\begin{abstract}
First quantum computers very recently have demonstrated “quantum supremacy” or “quantum advantage”: Executing a computation that would have been impossible on a classical machine. Today’s quantum computers follow the NISQ paradigm: They exhibit error rates that are much higher than in conventional electronics and have insufficient quantum resources to support powerful error correction protocols. This raises questions which relevant computations are within the reach of NISQ architectures. Several “NISQ-era algorithms” are assumed to match the specifics of such computers; for instance, variational optimisers are based on intertwining relatively short quantum and classical computations, thus maximizing the chances of success.
This paper will critically assess the promise and challenge of NISQ computing. What has this field achieved so far, what are we likely to achieve soon, where do we have to be skeptical and wait for the advent of larger-scale fully error-corrected architectures?
\end{abstract}

\section{Introduction}
In spite of classical computing technology being on an exponential trajectory since 1960s, quantum computing (QC) continues to fuel the fantasies of scientists and practitioners alike. QC promises asymptotic, and in some cases exponential, speedups for hard problems from domains such as cryptography, computational chemistry, simulation, or machine learning. Theoretical results on quantum algorithms have been complemented by a development of actual quantum hardware potentially capable of practically executing quantum computations, even though the demonstrated systems were of limited size and could be simulated on a classical computer within a meaningful time.

The ``quantum supremacy'' experiment \cite{14} showed computations on a 53-qubit (quantum bit) machine that the authors argued would be intractable on a classical compute server. Even though the latter argument is currently disputed \cite{27,44}, as other authors are proposing more efficient classical solutions for the problem of \cite{14}, this problem serves no practical or scientific purpose other than demonstrating ``quantum supremacy'' (or ``quantum advantage''). Using QC for solving a practical problem faster than any classical computer is still open at the time of writing.

One key challenge on the road to quantum advantage is increasing the number of qubits that a quantum computer can calculate on, as the asymptotic speedups naturally unfold their effects on larger problem instances. Less obvious but not less important, the \emph{quality} of the available qubits and operations on them (quantum gates) is essential for meaningful computations. Quantum states are fragile, and even error rates reported for the hardware in \cite{14} (which are among the best implementations that exist today) were orders of magnitude larger compared to conventional electronics. These error rates are expected to improve in the future, but not come close to the nearly error-free operation of classical computers.

When it comes to the near future of quantum computing given the non-trivial error rates of its hardware, two schools of thought exist with. The first is to deploy \emph{quantum error correction} (QEC), where several poor-quality physical qubits together constitute a \emph{logical qubit} with a more reasonable robustness. The idea is not unsimilar to processing encoded information in self-checking design; however, state-of-the-art QEC schemes \cite{38} have an overhead of hundreds or thousands physical qubits per logical qubit. The second approach, known under the name ``NISQ'', suggests to accept the inherent noise as given and try to find applications that can survive it without resorting to fully-fledged QEC. Some classes of algorithms are being investigated specifically with NISQ computers in mind~\cite{7}. Near-term NISQ computers are also the focus of this paper.

Since “noise” is an integral part of NISQ, understanding its origins, characterising its properties, quantifying its magnitude, and assessing its application-level impact are of utmost importance. To this end, this paper will be organised as follows. After some background information about QC in Section \ref{2}, Section \ref{8} will introduce the basic NISQ concepts and revisit the expectations raised since its inception. Section \ref{1} will discuss future developments in QC, going beyond NISQ.

Section \ref{4} will put today’s successful demonstrations of the NISQ principle for small instances of practically relevant problems into perspective. For example, certain computational chemistry tasks have been solved for small molecules, such as the hydrogen atom, for which classical solving methods work as well. What would be needed to extend such approaches to more complex tasks? Section \ref{3} will explain what role characterisation play in making NISQ computers useful. The approaches go far beyond the pass-fail tests used for conventional CMOS circuits. Individual qubits and ensembles of entangled qubits must be characterised using techniques such as randomised benchmarking~\cite{33}~or gate set tomography~\cite{16, 28}. Section \ref{7} will conclude the paper.

\section{Quantum Computing Background}
\label{2}

A quantum computer can perform computations on $n$ qubits, where one qubit might be, for instance, an ion, a photon, or a solid-state transmon circuit. An individual qubit can be set to two basis states $|0\rangle$ and $|1\rangle$, which correspond to two-dimensional column vectors $(1, 0)^{\rm T}$ and $(0, 1)^{\rm T}$, respectively. A single qubit assumes a state, which can be $|0\rangle$, $|1\rangle$, or their \emph{superposition} $\alpha_0 |0\rangle + \alpha_1 |1\rangle = (\alpha_0, \alpha_1)^{\rm T}$, where $\alpha_0$ and $\alpha_1$ are complex numbers and $|\alpha_0|^2 + |\alpha_1|^2 = 1$. An ensemble of $n$ qubits assumes a state described by a $2^n$-dimensional complex vector of \emph{amplitudes} $(\alpha_{0\ldots00}, \alpha_{0\ldots01}, \ldots \alpha_{1\ldots11})^{\rm T}$, where the indices of $\alpha$ are all n-bit combinations of 0s and 1s and again $|\alpha_{0\ldots00}|^2 + |\alpha_{0\ldots01}|^2 + \cdots + |\alpha_{1\ldots11}|^2 = 1$.

\begin{figure}
    \centering

    \begin{tabular}{@{}c@{}c@{}c@{}c@{}}
        \begin{tabular}{c}
            \hspace{3ex}Hadamard\\

            \hspace{3ex}\includegraphics[scale=1.35]{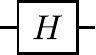}\\
            $\frac1{\sqrt2}\begin{pmatrix}1 & 1\\ 1 & -1\end{pmatrix}$
        \end{tabular}
         &
        \begin{tabular}{c}
            Pauli-X\\
            
            \includegraphics[scale=1.35]{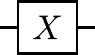}\\
            $\begin{pmatrix}0 & 1\\ 1 & 0\end{pmatrix}$
        \end{tabular}
         &
        \begin{tabular}{c}
            Pauli-Y\\
            
            \includegraphics[scale=1.35]{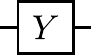}\\
            $\begin{pmatrix}0 & -i\\ i & 0\end{pmatrix}$
        \end{tabular}
         &
        \begin{tabular}{c}
            Pauli-Z\\
            
            \includegraphics[scale=1.35]{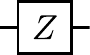}\\
            $\begin{pmatrix}1 & 0\\ 0 & -1\end{pmatrix}$
        \end{tabular}
    \\\\
    \end{tabular}
    
    \begin{tabular}{cc}
        \begin{tabular}{c}
            Controlled-NOT\\
            
            \includegraphics[scale=1.5]{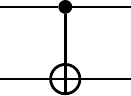}\\
            $\begin{pmatrix}1 & 0 & 0 & 0\\ 0 & 1 & 0 & 0\\ 0 & 0 & 0 & 1\\ 0 & 0 & 1 & 0\end{pmatrix}$
        \end{tabular}
         &
         
        \begin{tabular}{c}
            Example circuit:\vspace{1ex}\\

            \includegraphics[scale=1.35]{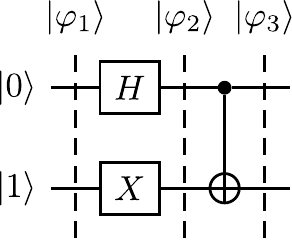}\\[0.25cm]
            \vspace{-6ex}
        \end{tabular}
    \end{tabular}

    \caption{Some important quantum gates and an example quantum circuit. 
}
    \label{5}
\vspace{-0.2cm}
\end{figure}

A \emph{quantum gate} $G$ acting on $k$ qubits is described by a $2^k \times 2^k$ complex \emph{unitary matrix} $U_G$. Examples of quantum gates (some of which play a key role in modeling errors) along with their symbols and matrices are shown in Fig.~\ref{5}. Note that all quantum gates are invertible and therefore have the same number of inputs and outputs. Gates acting in parallel on different qubits are combined using tensor product; gates acting sequentially can be combined using matrix multiplication. Fig.~\ref{5} includes an example circuit; let us follow its operation if the first and the qubit are initialised in states $|0\rangle$ and $|1\rangle$, respectively:
\begin{eqnarray*}
\ket{\varphi_1} & = & |01\rangle = \begin{pmatrix}1 \\ 0\end{pmatrix} \otimes \begin{pmatrix}0 \\ 1\end{pmatrix} = \begin{pmatrix}0 \\ 1 \\ 0 \\ 0 \end{pmatrix} \\
\ket{\varphi_2} & = & \frac1{\sqrt2}\begin{pmatrix}1 & 1\\ 1 & -1\end{pmatrix} \otimes \begin{pmatrix}0 & 1\\ 1 & 0\end{pmatrix} \ket{\varphi_1} \\
& = & \frac1{\sqrt2} \begin{pmatrix}0 & 1 & 0 & 1\\ 1 & 0 & 1 & 0\\ 0 & 1 & 0 & -1\\ 1 & 0 & -1 & 0\end{pmatrix}\begin{pmatrix}0 \\ 1 \\ 0 \\ 0 \end{pmatrix} = \begin{pmatrix}1/\sqrt2 \\ 0 \\ 1/\sqrt2 \\ 0 \end{pmatrix}\\
\ket{\varphi_3} & = & \begin{pmatrix}1 & 0 & 0 & 0\\ 0 & 1 & 0 & 0\\ 0 & 0 & 0 & 1\\ 0 & 0 & 1 & 0\end{pmatrix} \begin{pmatrix}1/\sqrt2 \\ 0 \\ 1/\sqrt2 \\ 0 \end{pmatrix} = \begin{pmatrix}1/\sqrt2 \\ 0 \\ 0 \\ 1/\sqrt2 \end{pmatrix}
\end{eqnarray*}
Note that state $\ket{\varphi_2}$ corresponds to the system being in superposition of $|00\rangle$ and $|10\rangle$, which can be attributed to its individual qubits: qubit 1 is in the superposition state $(0\rangle + |1\rangle) / \sqrt2$, and qubit 2 is in state $|0\rangle$ (the Pauli-X gate acts as an inverter). In contrast, state $\ket{\varphi_3}$ is a superposition of $|00\rangle$ and $|11\rangle$ and cannot be expressed by the individual qubits; this phenomenon is known as \emph{entanglement}.

It sounds quite unspectacular that most of a quantum computer's operations are multiplying matrices (of its gates) by the state vector. It seems a lot more spectacular that these vectors and matrices have an exponential size in the number $n$ of actual physical objects, i.e. qubits. One would expect radical speed-ups from a system that is seemingly able to hold $2^n$ intermediate results in the state vector $(\alpha_{0\ldots00}, \alpha_{0\ldots01}, \ldots \alpha_{1\ldots11})^{\rm T}$ and applying computations to all these results at once. The reality is more intricate: we cannot read out the values of the state vector directly, but can only perform a \emph{measurement} on it.

Measuring a qubit will result in either outcome $0$ or $1$.
If a qubit's state was $|0\rangle$ or $|1\rangle$, the measurement will deterministically yield  $0$ or $1$.
However, if a qubit is in a superposition $\alpha_0 |0\rangle + \alpha_1 |1\rangle = (\alpha_0, \alpha_1)^{\rm T}$, the outcome will be  $|0\rangle$ with probability $|\alpha_0|^2$ and $|0\rangle$ with probability $|\alpha_1|^2$. For example, measuring the second qubit in state $\ket{\varphi_2}$ in the example above will yield $|0\rangle$ with certainty, and measuring the first qubit will result in $|0\rangle$ or $|1\rangle$ with probability $|1/\sqrt2|^2 = 1/2$ each. The measurement is \emph{destructive}: after the measurement, the state of this qubit will change to the state corresponding to the measurement outcome, and all non-trivial information about $\alpha_1$ and $\alpha_2$ will be lost. An interesting effect happens in entangled states like $\ket{\varphi_3}$: measuring one qubit can determine the value of the other. For example, if measuring the first qubit in $\ket{\varphi_3} = (|00\rangle + |11\rangle) / \sqrt2$ yields $|0\rangle$ (this happens with 50\% probability), then the 2-qubit state collapses to $|00\rangle$; the next measurement of qubit 2 will result in $|0\rangle$ with certainty.

The destructive character of measurements prevents the seemingly obvious parallelism. It is \emph{not} possible to obtain the exact values $\alpha_j$ by copying the end-state of the circuit and measuring it multiple times. It \emph{is} possible to re-run the circuit multiple times from the beginning, but the required number of repetitions would offset any gain from the parallelism. Instead, quantum algorithms employ steps to transform (entangled) quantum states such that measurements yield useful information.

The description so far had assumed a perfectly working quantum computer. Real-world hardware is prone to disturbances due to noise, interaction with the environment, and imperfections of the physical apparatus~\cite{3}. One can distinguish different types of failures. \emph{Pauli noise} is modeled by Pauli-$X$, $Y$ and $Z$ gates (Fig.~\ref{5}) randomly added to different locations within the circuit. Somewhat counter-intuitive, this discrete fault model captures parametric (large and small) errors in the individual qubit states; this is because a qubit is measured at some point and even a small error either manifests itself (this is captured by Pauli errors) or not (then it has no effect).
Frequently used special cases of Pauli noise are \emph{depolarizing noise}, where Pauli-$X$, $Y$ or $Z$ gates are applied with uniform error rates independently to each qubit, and \emph{dephasing noise}, where only Pauli-$Z$ gates are randomly applied within the circuit.

Independent Pauli errors do \emph{not} cover correlated errors that can be generated by effects such as physical control line \emph{crosstalk} between qubits, but these effects can be bounded as joint Pauli errors occurring at the same time across the relevant qubits - the error is often bounded to weight two Pauli errors as the underlying physics of qubit systems is limited to pairwise Hamiltonian interactions. 
\emph{Quantum error correction} (QEC) protocols make use of heavily entangled states to form highly redundant ``logical qubits'', such that errors affecting one or few physical qubits are compensated by error-unaffected redundant physical qubits within the same entangled state. However, NISQ computers follow a different route.

\section{Defining NISQ, and How Useful Will the NISQ Era Be?}
\label{8}
\indent In terms of quantum hardware, the necessary components to build a scalable quantum computer (in a variety of different systems, including superconductors) are well known \cite{46,53,43,31}.  At a minimum, a hardware device requires a regular array of qubits arranged in a square 2D grid.  Each qubit needs the ability to initialise in a known state, perform arbitrary single qubit operations, be able to be measured and be able to couple to the four neighbouring qubits in the array.  All of these fundamental operations need to occur with an error rate of 0.1\% or lower \cite{38}.  This is the bare minimum for a scalable system.  

As anticipated by many, the first realisation of quantum computing technology has occurred over the cloud, with users logging onto dedicated hardware over the classical internet. These types of `quantum in the cloud’ systems began with the connection of a two-qubit photonic chip to the classical internet by the University of Bristol in 2013 and accelerated significantly in 2016 with the introduction of IBM of their quantum experience platform.  We now see both free and paid services offered by IBM, Microsoft, Amazon, Xanadu and Rigetti, across a variety of hardware modalities for small scale quantum computing chipsets up to 65 physical qubits.  This has spurred the so-called era of noisy intermediate-scale quantum (NISQ) \cite{29} research.

Noisy intermediate-scale quantum also refers to quantum algorithms that are small enough to be faithfully executed on near term, low qubit count, high error rate, quantum hardware without the need for resource-costly error correction protocols.   While NISQ algorithms exist in principle---quantum supremacy is a quintessential example \cite{14}---an added caveat is that the algorithm needs to be either scientifically and/or commercially viable---quantum supremacy is not.  Consequently NISQ algorithms must be highly compact but still either reach the quantum supremacy regime---where the problem simply cannot be solved on any classical computer---or reach the regime where it is more cost efficient---either in terms of actual dollars or in terms of computational time---to run the algorithm on a quantum computer.  NISQ algorithms satisfying the commercial or scientific viability condition are under active research, but do not currently exist.

The actual definition of NISQ computation is somewhat nebulous and depends on who you ask.  However, to a first approximation we can examine NISQ compatibility by calculating for a given algorithm the value $A = n\cdot d$, called \emph{area}, where $n$ is the number of qubits the algorithm needs and $d$ is the number of time-steps in the algorithm.  If the physical error rate of the qubits and gates in a physical quantum computing chip is $p < 1/A$, the algorithm is likely NISQ compatible.  This is due to the error sensitivity of quantum algorithms.  A quantum computation is akin to an optical interferometer and its ability to perform efficient computation comes about through a delicate interference effect of the computational wavefunction as gate operations are applied.  As with physical interferometers, where a minor misalignment can completely disrupt the interference effect and make the device non-functional, even one error in a quantum algorithm can disrupt the required computational interference effect, resulting in incorrect output.  It has been demonstrated for a variety of quantum algorithms that single errors can dramatically reduce the probability of success of generating the correct output~\cite{17, 5}.
This makes $p < 1/A$---which represents the error rate needed such that one error occurs during execution on average---a good bound for the error tolerance of an algorithm.

The difficulty is finding a quantum algorithm that is comfortably in the regime where classical algorithms simply cannot be used at the same time as still being small enough to satisfy the $p<1/A$ bound which defines when error correction or other mitigation techniques are needed.  The quantum supremacy formalism that was used by the Google Sycamore processor \cite{14} are, by definition, the smallest quantum circuits known that are provably difficult to simulate on a classical computer.  They were constructed explicitly for this purpose---this is why they are not practically useful beyond demonstrations of quantum supremacy.  The Google team simulated a 53-qubit circuit over a depth of approximately 40, giving $A \approx 2100$.  Even this circuit was `border-line' in terms of being implementable on classical hardware.  A preprint paper in early 2021 has in fact presented results that replicated the Google experiment using new techniques in classical algorithms based on tensor contraction\cite{44}.  

A circuit with $A \approx 2100$ would therefore require physical errors in the quantum processor of $p <  5\times 10^{-4}$ to occasionally run without errors.  Not even the Google Sycamore processor had this level of accuracy.  Single and two-qubit gates in the experiment had, on average, error rates of approximately 0.1-1\% while measurement gates had errors up to 5\% for simultaneous readout \cite{14}. 

Hardware errors for a variety of quantum processors have decreased from of order 10-50\% when basic qubit operations were first demonstrated in the late 1990's and early 2000's to, of order, 1-0.1\% today (over two decades later).  
This is extremely impressive from an experimental point of view, and the ability to replicate the fabrication of low error rate quantum chips is getting better and better.  However, given that the smallest quantum circuit that has provable quantum advantage---that is commercially and scientifically useless---requires physical error rates at least another two orders of magnitude lower to be unambiguously achievable in a quantum processor, it is difficult to accept the prospect that an algorithm that can be implemented with higher physical error rate that can simultaneously revolutionise a scientific discipline or a commercial industry is possible.  The only other option is that experimental progress in reducing physical error rates in chip-sets accelerates rapidly, such that the next two orders of magnitude in error improvement occurs over the next two or three {\em years} rather than the next two or three {\em decades}.

Consequently, NISQ currently has no identified, commercially relevant application at the 50-1000 qubit level.  There are also good reasons to believe that finding such applications (in an era where error rates on quantum chips will be of the order of 0.1\% -- 1\% at best) will not be possible.  However, there are no results that prove that the existence of these algorithms are impossible, so there is still strong motivation to keep searching. This search should be done in a way that addresses these fundamental issues rather than sidestepping them, or in some unfortunate cases, simply obfuscating the problems.

\section{What Happens Beyond NISQ?}
\label{1}

Beyond the NISQ era is one of two possibilities.  Either physical error rates in quantum computing chips are drastically reduced---via a combination of better fabrication physics, better quantum control and passive error mitigation techniques \cite{36}---or active techniques are employed to reduce error rates which fall under the umbrella of Quantum Error Correction (QEC)~\cite{3}. 

QEC uses redundant qubit encoding, active measurement and feedback to stabilise quantum information over long periods of computational time.  It is a highly successful theoretical discipline and is unarguably one of the main pillars of quantum information science.  QEC is commonly combined with the method of fault-tolerant circuit design, where error correction protocols and logic operations are performed in a manner such that physical errors do not `cascade' out of control.  The combination of these two techniques lead to one of the most important results in quantum computing, the threshold theorem.

The threshold theorem for fault-tolerant quantum computing \cite{48} states that with a polylogarithmic overhead of physical resources---qubits and time---an arbitrarily long quantum circuit can be implemented provided the physical errors of the hardware are less than a critical value, $p_{th}$, known as the fault-tolerant threshold. Currently, the threshold for the most commonly used QEC scheme for large-scale architecture design, the surface code, is $p_{th} \approx 0.6\%$.  This means that several hardware platforms already have reached physical error rates satisfying the fault-tolerant threshold for QEC in a subset of operations, although it should be noted that no hardware system has yet demonstrated error rates below threshold for a universal set of gate operations on a single device. 

The unfortunate issue with QEC is the drastic increase in physical qubit resources once it is even implemented to a minimal degree.  Once error correction is incorporated into an algorithm, the total number of physical qubits required in the quantum chipset can easily jump from 50-100 qubits to 50,000-100,000.  This includes both the raw overhead in performing the qubit encoding and ancillary physical resources to maintain fault-tolerance in operational logic. 
This would only be for a small amount of error correction, enough to reduce error rates from approximately 0.1\% on the physical level to the order of 0.0001\% at the encoded level.
This number of physical qubits is just not possible to engineer in the near future, in any hardware system.  

The error correction overhead for large-scale algorithm, such as Shor's factoring algorithm or quantum simulation is very high.  However, two points should be strongly emphasised. 
\begin{itemize}
    \item Factoring or quantum simulation are extremely large algorithms, even if they scale polynomially with problem size.  
    
    Given the error sensitivity of quantum algorithms, this means that the effective error rate required for encoded qubits can be of the order of $10^{-13}\%$ or lower.  
    Hence, QEC needs to provide at least 12 orders of magnitude of error improvement.  This massive amount of suppression does not come for free.
    \item Classical techniques for circuit optimisation both at the algorithmic level and at the QEC level has already done an extraordinary job at reducing the overhead.  For example, in 2012, an explicitly compiled, error-corrected analysis of Shor's algorithm estimated that of the order of 100 billion qubits would be needed to break RSA-2048 encryption \cite{26}. By 2019, this number had been reduced to 20 million \cite{39}, a reduction by a factor of 5,000 without changing any of the operating assumptions of the underlying hardware.
\end{itemize}
Ultimately, the demands required on hardware accuracy for large-scale algorithms will necessitate active error correction in the long term.  How far experimental improvements in fabrication, control and passive error mitigation can go before active techniques need to be implemented is up for debate, but there will be (and may already be) decisions made to the most cost effective and expedient way to move in the future.  Either invest in reducing physical error rates by orders of magnitude or invest in being able to make chip-sets containing very large numbers of integrated qubits quickly and cheaply.

\section{Analysing NISQ Compatibility of Algorithms}

\label{4}
Recent progress in quantum computing technologies~\cite{14, 11, 7, 13}~offers quantum computational resources that have not been available before.
Although only noisy and intermediate-sized quantum computational resources can be exploited by current NISQ computers, recent experiments yield results that are challenging to obtain using classical resources~\cite{14, 13}.
In the light of these experiments, increasing quantum resources of NISQ computers and growing business adaptation, it is crucial to determine the resource requirements of a computation.
Assessing these requirements determines the set of suitable NISQ computers, can help to improve the implementation of a computation, and informs whether quantum error correction is necessary.

The resource requirements of a computation consist of the number of qubits, the depth
and the tolerable error rate.
The first two requirements can be computed by inspection~\cite{15}.
The tolerable error rate can be determined through approaches ranging from guidelines on the size of the quantum computation, simulations subject to errors, and experiments on quantum computers to reliability models of quantum computers based on machine learning.
The determination of the tolerable error rate must be able to consider noise sources ubiquitous in NISQ computers, arbitrary quantum computations, flexible definitions of success and must be able to make statements about a wide range of quantum computing technologies.

In~\cite{29}~and~\cite{10}~bounds on the quantum computation size and error rate $p$ for successful computations are described.
The work in~\cite{29}~states that the error rate may not be much larger than $G^{-1}$, with $G$ denoting the number of quantum gates constituting the quantum computation.
The work in~\cite{10}~states the observation
\begin{equation}
p < (n\cdot d)^{-1}
\end{equation}
with $n$ qubits used in a computation that has depth $d$. These guidelines provide a rough estimation and are not suitable to distinguish between different success criteria or quantum computations that have the same size but exhibit a different error behaviour.

A more accurate approach to determine the tolerable error rate of a quantum computation is to conduct simulations subject to errors~\cite{17, 1, 22, 12, 5, 23, 21, 24, 8}. Using Monte Carlo or exhaustive error simulations, the error rate at which the quantum computation starts failing the target success criteria can be determined. Here, the employed error model significantly contributes to the accuracy, simulation effort and generality of the determined tolerable error rate. If an error model is employed that is fitted to the dynamic adverse physical processes (noise processes) in a specific quantum computer, the determined error rate requirement can often not be generalized to other quantum computers~\cite{2}~or quantum computing technologies. Alternatively, if the employed error model is too general, it may not represent noise processes sufficiently well and thus lead to incorrect predictions. While simulations subject to errors are a suitable choice to determine the tolerable error rate, they also incur an exponential runtime and space overhead in the number of qubits in general. It is therefore crucial to reduce the simulation effort by a suitable choice of error model and simulation techniques~\cite{17}.

Another option is to probe the success probability of a quantum computation on a target quantum computer. 
While this can be performed efficiently, it is challenging to generalise the results to other quantum computers of the same generation due to the dynamic error rates of NISQ computers, and the results are not valid for other quantum computing technologies. 
In addition, a computation is performed on a quantum computer with respect to a specific error rate that can only be increased in a limited way~\cite{0}. Thus, the error rate requirement of a quantum computation can not be determined in general since only a small range of error rates can be probed.

Reliability models for quantum computers based on machine learning have also been proposed~\cite{40}. These models are trained on a specific quantum computer and uses features such as quantum computation size or structure, auxiliary experiments and success probabilities of previous quantum computations to yield a prediction about the success probability of future quantum computations. While these methods exhibit a high prediction accuracy for some quantum computers and computations, these approaches have not been demonstrated to generalise well over multiple quantum computers or different quantum computing technologies.

Thus, for determining the error rate requirement of a quantum computation in the NISQ era, simulations subject to errors is a suitable choice. Arbitrary noise sources, quantum computations and success criteria can be considered with simulations. While the runtime overhead is exponential in the number of qubits, there currently is only a small number of publicly available NISQ computers that can not be simulated in reasonable time.
We will now give details about commonly used success criteria, simulation methods, error modeling and results for a set of small-scale quantum algorithms.

\subsection{Success Criteria}
The success of a quantum circuit computation is determined depending on the available reference information of the ideal result.
In the simplest case, when the target state is known and exact probability amplitudes are available upon simulation, the fidelity measure  $F(\ket{\psi_t}, \ket{\psi_e})$ is often used to quantify how much an erroneous state $\ket{\psi_e}$ deviates from a target state $\ket{\psi_t}$~\cite{18}.
However, this is often not applicable as quantum circuit computations on a quantum computer only return measurement results and the target state is often not known.
Therefore, a number of success criteria were proposed and used in previous works such as~\cite{6}:
\begin{itemize}
    \item The measurement probability of the correct result in single outcome quantum circuits is above a certain threshold.
    \item Measurements are in a set of acceptable results with high probability larger than a threshold. Such a set can e.g. be defined by the Hamming distance or the heavy output~\cite{20}.
    \item The deviation between the measurement output distribution and the expected distribution is smaller than a threshold according to some measure of distance~\cite{6}.
    \item The measurement result probabilities are consistent with predictions, i.e. the cross-entropy of results is low~\cite{37}.
\end{itemize}
Since quantum circuit computations on a quantum computer are subject to errors and sampling noise, above probabilities can not be determined with certainty~\cite{6}. It is therefore essential to also consider a metric of confidence for reaching a defined success criteria. In addition, applying a success criterion to a quantum computation may not make the results classically tractable.

\subsection{Simulation Methods}
Simulating a quantum circuit incurs an exponential runtime and space overhead in general~\cite{18}.
However, specific quantum circuit structures and properties can be exploited by simulators based on tensor networks~\cite{9}, matrix product states~\cite{47}, decision diagrams~\cite{52}~or stabilizers~\cite{4, 19}.
These simulators exhibit a better runtime or memory requirement for specific quantum circuits than general simulators based on the state vector or density matrix.

In general, simulating a $n$-qubit state requires a state vector to store and manipulate $2^n$ complex amplitudes whereas the density matrix simulator has to store and manipulate $2^n$ $n$-qubit states. Simulating 53-qubit quantum circuits such as the quantum circuit in Google's quantum supremacy experiment using a state vector or density matrix simulator imposes a large memory requirement: $72.1$ petabytes and $6.49 \cdot 10^{17}$ petabytes would be required for the state vector and density matrix simulator respectively, if one complex number is stored using two double-precision floating-point numbers.

\subsection{Error Modeling}
Noise in NISQ computers is ubiquitous and affect the computation of a quantum circuit in various ways.
In general, a qubit can be subject to coherent or incoherent noise, be lost, leak out of the computational space, or be erroneously initialised or read out~\cite{3}.
These effects can be modeled by \emph{device-oriented} or \emph{device-agnostic} error models.
In a device-oriented error model, the adverse physical processes in a quantum computing device are replicated to model the errors affecting a computation.
Device-agnostic error models, however, consist of a set of operators that cover the effects of relevant noise sources.
The choice of error model has consequences for the prediction accuracy, applicability of results, quantifiability of the model, and simulation effort of simulations subject to these error models.

A device-oriented error model fitted to one specific quantum computer would be expected to yield the best prediction accuracy. However, the results of simulations subject to that error model would not be applicable to different quantum computers or quantum computing technologies. 
Furthermore, estimating each error parameter of such an error model through characterisation protocols can be complex, as described in more detail in Section \ref{3}.
Device-oriented error models also incur a large simulation effort since they require the density matrix formalism in general.

A device-agnostic error model is applicable to diverse quantum computers, if the noise parameters are quantified on them using characterisation protocols. Thus, if it is known under which noise parameters the success criterion of a quantum circuit computation is met, we can determine suitable NISQ computers that do not exceed the determined noise parameter. Recently, quantum supremacy experiments confirmed that the device-agnostic Pauli error model is sufficiently accurate to make predictions about the success probability~\cite{14}. The Pauli error model also only incurs a low simulation effort for arbitrary quantum circuits since the error operators constituting the error model are in the Clifford set and can be simulated using the state vector or stabilizer simulator.

\subsection{Results}
In this section, we show results on various quantum circuit with up to 16 qubits and 214 gates subject to the Pauli error model with uniform error rates (depolarizing noise).
The tolerable Pauli error rate given a success probability of $66\%$ and the success probability given a Pauli error rate of $0.15\%$ are reported.
The tolerable Pauli error rate specifies the error rate at which a quantum computation can still be performed successfully given a target success criterion and probability.
The Pauli error rate of $0.15\%$ was chosen as this is currently the lowest (single-qubit gate) error rate exhibited by one of the largest NISQ computer~\cite{14}.
The evaluated quantum circuits implement the Grover algorithm~\cite{42}, arithmetic functions~\cite{50}, Bernstein-Vazirani (BV) algorithm~\cite{54}~(using CNOTs), the quantum Fourier transform (QFT)~\cite{51}~and the hidden linear function (HLF)~\cite{49}.
Furthermore, and quantum circuits for chemistry applications (RYRZ~\cite{7}, UCCSD~\cite{45}) were evaluated.

The evaluated quantum circuits were generated using Qiskit \cite{41} for a general quantum computer with single-qubit rotations, the controlled-NOT two-qubit gate and without geometric constraints.
Geometric constraints imposed by many quantum computing technologies such superconducting qubits~\cite{14} decrease the herein reported tolerable Pauli error rate and success probability since satisfying these constraints require the insertion of additional error-prone operations.

The evaluated success criterion was the measurement probability of the correct result for single outcome algorithms such as BV, 
and fidelity for all other.

The quantum circuits were simulated subject to the Pauli error model using a state vector simulator~\cite{5}.
For combinations of quantum circuits and Pauli error rate where less than one Pauli error occurs during the quantum circuit computation on average, an exhaustive error simulation was employed.
For this exhaustive error simulation, one Pauli X, Z, or Y error was simulated successively at every space-time location in the quantum circuit and the impact of these errors on the target success criterion was scaled according to the specified Pauli error rates~\cite{17}.
For combinations of quantum circuits and Pauli error rates that incur more than one Pauli error per quantum circuit computation on average, Monte Carlo simulations were conducted to obtain the success probability and tolerable Pauli error rate.

\subsubsection{Success Probability}
\begin{figure}[t!]
  \centering
  
  \includegraphics[width=\linewidth]{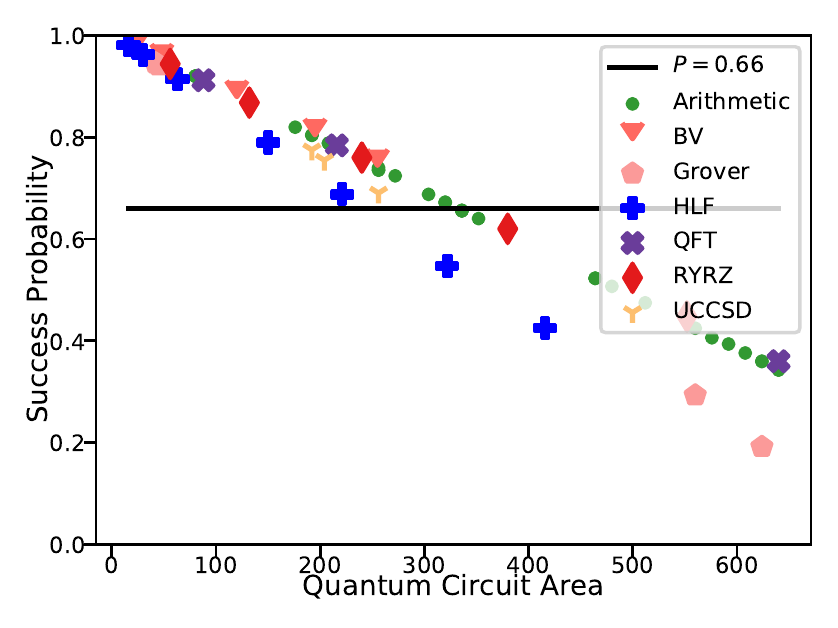}
  
  \caption{Success probability of quantum circuits subject to a Pauli error rate of 0.15\% and no geometric constraints. \label{0}}
  \vspace{-3ex}
\end{figure}

Fig.~\ref{0}~shows the success probability of the evaluated quantum circuits at a Pauli error rate of $0.15\%$.
On the x-axis, the area of the evaluated quantum circuits is shown.
All quantum circuits with an area smaller than $322$ could be executed with a success probability of at least 66\%.

All evaluated BV and UCCSD quantum circuits could be computed successfully.
For other quantum circuits and moderate quantum circuit sizes, the probability of successfully computing the quantum circuit quickly falls below 66\%.
These results show that the quantum circuit area is a good indicator for determining the success probability for many of the evaluated quantum circuits.
\subsubsection{Tolerable Pauli Error Rate}
\begin{figure}[t!]
  \centering
  
  \includegraphics[width=\linewidth]{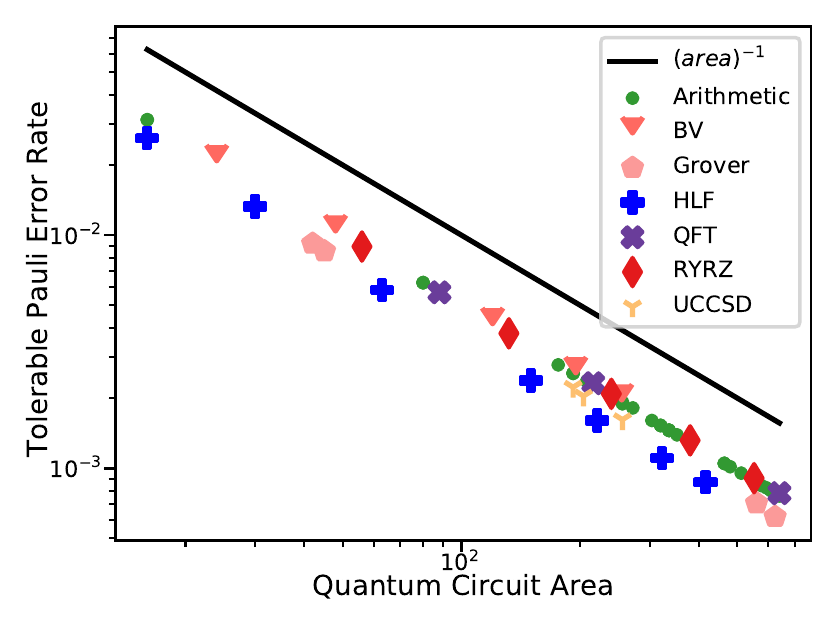}

  \caption{Tolerable Pauli error rate of the evaluated quantum circuits for a success probability of 66\%.\label{6}}
   \vspace{-3.6ex}
   
\end{figure}

Fig.~\ref{6}~depicts the tolerable Pauli error rate of the evaluated quantum circuits for a success probability of 66\%. The x-axis shows the quantum circuit area. Both axes are in log-scale. The bound of the error rate on the quantum circuit area $(n\cdot d)^{-1}$ stated in the literature~\cite{10}~is also shown as a black line.

All evaluated quantum circuits only tolerate a Pauli error rate that is smaller than the inverse of their quantum circuit area.
There are small difference in the tolerable Pauli error rate between quantum circuits implementing different quantum algorithms.
Quantum circuits for arithmetic functions tolerate a Pauli error rate that is mostly the average of evaluated quantum circuits with the same quantum circuit area.
BV, RYRZ and QFT quantum circuits tolerate a slightly larger Pauli error rate and HLF, UCCSD and Grover quantum circuits only tolerate a slightly smaller Pauli error rate.
The tolerable Pauli error rate of HLF quantum circuits constitute a lower bound for the evaluated quantum circuits.

These results confirm the bound given in~\cite{10}~and indicate that the evaluated quantum circuits do not tolerate one Pauli error on average.
The tolerable Pauli error rate is mainly dominated by the area of a quantum circuit;
large difference due the structure of a quantum circuit or other properties were not observed.

\section{Characterising the underlying hardware}
\label{3}

Above, we have seen the important impact of physical error rates onto the applicability of quantum error correction protocols and NISQ algorithms. In this section, we will therefore present some of the most commonly used methods for actually measuring the error rates. These methods do not require any knowledge about how the quantum gates are physically realised. Instead, the basic idea is the following: we execute suitably chosen quantum circuits on the given hardware and count the measurement outcomes.
The desired information concerning the errors is then extracted by comparing the measured outcomes with the outcomes expected in the absence of errors.

\subsection{Modeling Errors in Quantum Circuits}

To introduce these methods, we first explain some basic concepts concerning the mathematical modelling of errors.
In section~\ref{2}, we have introduced the state of a qubit as a two-dimensional complex vector $|\phi\rangle=\left(\alpha_0, \alpha_1\right)^{\rm T}$. Such a state is called a \emph{pure quantum state}. More generally, however---and especially so in the presence of errors---the qubit may be in a \emph{mixed state}, which is described by a $2\times 2$ \emph{density matrix}:
$$\rho = \left(\begin{array}{cc}\rho_{00} & \rho_{01} \\ \rho_{10} & \rho_{11}\end{array}\right)$$
with $\rho=\rho^\dagger$ (self-adjoint; the symbol $\dagger$ denotes transposition and complex conjugation), ${\rm Tr}(\rho)=1$ (normalised) and $\rho\geq 0$ (positive semi-definite).
In the special case of a pure state $\phi$, we have $$\rho=|\phi\rangle\langle\phi|=\left(\begin{array}{cc}
|\alpha_0|^2 & \alpha_0\alpha_1^* \\ \alpha_0^*\alpha_1 & |\alpha_1|^2 \end{array}\right),$$
where $\langle\phi| = (\alpha_0^*,\alpha_1^*)$.
In a similar way, states of $n$ qubits are described by $2^n \times 2^n$ dimensional density matrices.
 
If a unitary operator (or quantum gate) $U$ is applied to the qubits, their density matrix changes according to 
$$\rho' = {\mathcal E}_U(\rho):=U \rho U^\dagger,\ \ \text{where}\ U^\dagger U ={\mathbbm 1}$$
If $\rho$ is a pure state, then $\rho'$ is also a pure state. For this reason, there is no need to consider density matrices if the quantum computer exhibits no errors and all gates are perfectly unitary. In the presence of errors, however, the resulting quantum operation assumes the following general form (completely positive, trace preserving map) \cite{30}:
$$\rho' = {\mathcal E}(\rho)=\sum_j A_j \rho A_j^\dagger,\ \ \text{where}\sum_j A_j^\dagger A_j = \mathbbm 1$$
This can be interpreted as follows: instead of a single well-defined unitary operation $U$, one of several possible operators $A_j$ is applied to the qubits with probability
$p_j={\rm tr}(A_j\rho A_j^\dagger)$. 
As an example, the depolarising channel for a single qubit is defined by $A_0=\sqrt{1-p}~ \mathbbm 1$, $A_1 = \sqrt{\frac{p}{3}} X$,  $A_2 = \sqrt{\frac{p}{3}} Y$ and $A_3 = \sqrt{\frac{p}{3}} Y$. In other words, with probability $1-p$ (in this case independent of $\rho$), the state $\rho$ remains unchanged (no error), whereas an error amounting to the application of Pauli-$X$, $Y$ or $Z$ occurs with probability $p/3$ each.

It can be shown \cite{30} that the  sum over $j$ in the above general representation of a quantum operation  can be restricted such as to contain at most $4^n$ different terms. Thereby, a noisy quantum operation is defined by up to $4^n$ different $2^n \times 2^n$ matrices $A_i$ (called \emph{Kraus operators}), or---equivalently---by a single $4^n \times 4^n$ matrix ${\mathcal E}$
(sometimes called \emph{superoperator}).

There are several ways for defining the \emph{error rate} of a quantum operation.
 Suppose we want to implement the unitary quantum gate $U$, but realise $\mathcal E$ instead. Then,
we define the corresponding error operator by  $\Lambda=({\mathcal E}_U)^{-1}\circ \mathcal{E}$ (i.e. the inverse of the desired operation ${\mathcal E}_U$ is applied after $\mathcal E$, such that $\Lambda={\mathbbm 1}$ in the absence of errors). For a given initial pure state $\rho =|\phi\rangle\langle\phi|$), we can determine the fidelity $F(\phi)=\langle\phi|\,\Lambda(\rho)\,|\phi\rangle$ quantifying how well the state is preserved in spite of the error. A uniform average over all initial states yields the average fidelity $\overline F= \int {\rm d}\phi~F(\phi)$ with corresponding {\em average error rate} 
$$r = 1 -\overline{F}$$
As an example, in case of the $n$-qubit depolarising channel
$$\Lambda_{\rm dep}(\rho)=(1-p) \rho + \frac{p}{2^n}\mathbbm{1}$$
the average error rate $r$ is related to the 
Pauli error rate
$p$ by
$$r=\left(1-\frac{1}{2^n}\right)p$$
The difference between $r$ and $p$ can be traced back to the fact that, even if the state is perturbed by the undesired application of a Pauli operator, it may still retain some overlap with the original state. Apart from the average error rate, also other error measures are used, e.g. the diamond norm in connection with quantum error correction threshold theorems~\cite{48}. 

\subsection{Randomised Benchmarking}

A robust and scalable way for measuring the average error rate is provided by the method of {\em randomised benchmarking} \cite{34,33}. Here, random sequences of quantum gates are executed on the noisy quantum hardware and the  measurement results are compared to the ideal result expected in the absence of errors.
This method is proven to be scalable (i.e. applicable for a large number of qubits), which is a remarkable property given the fact that, in general, quantum circuits with a large number of qubits cannot be simulated on classical computers. In randomised benchmarking, however, the quantum circuits are chosen to consist of so-called \emph{Clifford gates}, which \emph{can} be efficiently simulated classically (according to the Gottesman-Knill theorem \cite{30}). This makes it possible to determine the ideal (i.e. error-free) result as a benchmark.

These Clifford gates form a finite group
$\{C_1,C_2,\dots,C_{|{\rm Clif}|_n}\}$
 (of size $|{\rm Clif}|_n=2^{2n} 2^{n^2}\prod_{j=1}^n(2^{2j}-1)$ for $n$ qubits) generated by the single-qubit Hadamard gate $H$, the two-qubit controlled NOT-gate (see Fig.~\ref{5}) and the single-qubit phase gate $S=\left(\begin{array}{cc} 1 & 0 \\ 0 & i\end{array}\right)$. Each element of the Clifford group can be decomposed into $O(n^2)$ of these elementary gates. Moreover, a \emph{random twirl over the Clifford group} turns any quantum operation $\Lambda$ into the depolarising channel
$$\Lambda_{\rm dep}=\frac{1}{|{\rm Clif}|_n}\sum_{j=1}^{|{\rm Clif}|_n} C_j^\dagger \circ \Lambda \circ C_j$$
with the same average error rate $r$. 
This makes it possible to characterise the error of $\Lambda$ by a single number $r$: the average error per Clifford gate.

\begin{figure}
	\includegraphics[width=8.6cm]{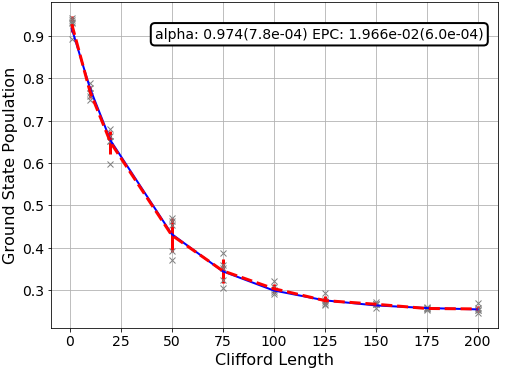}
	\caption{Result of a two-qubit randomised benchmarking experiment performed on the recently installed Fraunhofer-IBM quantum computer \lq ibmq\_ehningen\rq. 
	The average error rate per Clifford gate, extracted from an exponential fit (blue line) to experimental data (red dashed line, obtained as mean value over 5 different sequences of Clifford gates with various lengths) results as $r=0.01966$.
	Since the dominant error originates from CNOT gates (rather than from the single-qubits gates $H$ and $S$), and the average number of CNOT gates per Clifford gate used in this experiment is $1.485$, the resulting error rate per CNOT gate turns out as $0.01966/1.48=0.0132$.\label{9}}
\end{figure}

More precisely, the randomised benchmarking protocol works as follows: the quantum computer starts in the standard initial state $|00\dots 0\rangle$ (all qubits in state $|0\rangle$). Then, a sequence $C_{k_m}\dots C_{k_2}C_{k_1}$ of $m$ random Clifford gates (uniformly chosen from the above group)
 is applied. At the end, another Clifford gate $C_{k_{m+1}}=(C_{k_m}\dots C_{k_2}C_{k_1})^{-1}$ is applied, which---in the absence of errors---reverses the effect of the previous sequence. Therefore, the expected ideal measurement results is $00\dots 0$. With errors, however, the probability $p_{00\dots 0}$ to measure this state will be smaller than 1. Repeating this procedure $N$ times with $k=1..N$ for various lengths $m$, the measured results are fitted according to
$$p_{00\dots 0}=A_0 (1-p)^m +B_0$$
see Fig.~\ref{9}.
This fit determines the 
Pauli error rate
$p$ of the corresponding depolarising channel,
from which the average error per Clifford gate results 
in
$r=\left(1-\frac{1}{2^n}\right)p$.
The remaining parameters $A_0$ and $B_0$ take into account so-called \emph{SPAM errors}, i.e. errors in the preparation of the initial state and the measurement of the final state, as well as an edge effect from the error on the final gate. The above fitting formula is strictly valid in the case where the errors $\Lambda$ of each gate are identical and time-independent. However, a weak dependence of the errors on gates and on time can be included, resulting in a slightly more complicated fitting formula. 
Moreover, a modification of the above described protocol, called \emph{interleaved randomised benchmarking} \cite{25} can be used to determine different error rates for individual Clifford gates.

\subsection{Gate Set Tomography (GST)}

As already mentioned above, the main advantage of randomised benchmarking is its efficiency and scalability with respect to the number of qubits. On the other hand, it only yields a rough characterisation of errors in terms of a single number (the average error rate). Although this provides  useful information concerning the quality of the quantum hardware, it would be desirable to get a more detailed description. The average error rate only tells us how frequently errors occur, but nothing about what kind of errors these are. More detailed knowledge can be useful, e.g. to devise specific quantum error correction schemes that correct certain errors better than others, possibly at lower cost. 

Such a detailed characterisation is provided by the method of \emph{gate set tomography} (GST) \cite{16,28}. Here, we consider a certain set of
quantum gates $\{G_1,G_2,\dots\}$ and seek to determine their actual implementation on the noisy quantum hardware. For this purpose, the effect of each gate is modelled as a general quantum operation, i.e. each gate $G_i$ corresponds to a $4^n\times 4^n$ dimensional superoperator (see above).
In general, the complete characterisation of a quantum operation (known as
\emph{quantum process tomography} \cite{32}) requires the preparation of at least 
$4^n$ different (linearly independent) initial states, subsequent application of the respective quantum operation on all these states and finally, measurements with respect to $4^n$ different final states. When working with a gate-based quantum computer, however, we usually have only one initial state at our disposal (e.g. all qubits in state $|0\rangle$), and measurements are only performed with respect to the standard basis (i.e. the outcome corresponds to state  $|0\rangle$ or $|1\rangle$ for each qubit). Although the required complete set of initial and final states can be generated by applying suitable quantum gates (e.g. a Hadamard gate to generate the superposition $\frac{1}{\sqrt{2}}(|0\rangle+|1\rangle)$ as initial state), these gates themselves suffer from errors, which have to be distinguished from the errors of the gate one seeks to characterise. Apart from that, also the preparation of the initial state $|00\dots 0\rangle$ as well as the final measurement with respect to the standard basis exhibit errors (SPAM errors) that must be taken into account.

The solution provided by GST is to characterise a whole set of gates simultaneously and self-consistently.
For this purpose, a large number of quantum circuits, each consisting of a specific sequence of gates chosen from the respective gate set, are executed on the noisy quantum hardware. Then, the quantum operation corresponding to each gate, as well as the SPAM errors mentioned above, are extracted from the measurement results (More precisely, gates and SPAM errors can only be characterised up to a global gauge transformation \cite{28} which, however, is irrelevant for practical purposes, since it does not affect the observable outcomes of any quantum circuit). In the simplest implementation of gate set tomography (LGST---linear gate set tomography), this extraction can be performed using methods from linear algebra (e.g. matrix inversion), whereas numerically  more expensive techniques like maximum likelihood estimation are required in more advanced implementations  that yield more accurate results (long-sequence GST) \cite{28}. Due to the exponentially large number of parameters that have to be estimated ($4^n\times 4^n$ for each quantum gate), however,
a complete characterisation of errors as performed by GST is realistically feasible only for a small number of qubits (1 or 2, in practice).

In summary, we have discussed two of the most commonly used methods for characterising quantum hardware: randomised benchmarking, which can be applied for many qubits, but yields only a rough characterisation of errors, and gate set tomography, which provides a detailed characterisation for a few qubits. An active area of research consists of finding intermediate methods, which are scalable, but yield more detailed information than randomised benchmarking. A promising idea in this direction is to restrict the range of possible correlations (or crosstalk) between the errors of gates applied to different qubits by modelling the errors as a Gibbs random field. This approach has recently been used  to characterise errors in a 14-qubit quantum computer \cite{35}.

\section{Conclusions}
\label{7}

Quantum computing promises spectacular breakthroughs, enabling computations that were deemed impossible by classical computers. This capability stems from fully utilising properties of physical objects that combine an unprecedented sophistication with an extreme fragility. Understanding and controlling errors in quantum circuits is the holy grail on the road to achieving quantum advantage for practically useful problems. In this paper, we attempted a realistic view on the current progress of this journey. We critically reviewed the NISQ paradigm and identified its realistic capabilities but also its obvious limitations. We provided an overview about methods for both: analysing a quantum algorithm using the simplified, device-agnostic Pauli error model, and establishing far more detailed knowledge about the error mechanisms of a given quantum machine by means of randomised benchmarking and gate set tomography.

\balance

\renewcommand{\baselinestretch}{1.01}
\small\normalsize
\section*{Acknowledgments}
Parts of this work are supported by project QORA within the Competence Center Quantum Computing Baden-W\"urttemberg (funded by the  Ministerium f\"ur Wirtschaft, Arbeit und Wohnungbau Baden-W\"urttemberg), and by a project funded by Carl Zeiss foundation.

\renewcommand{\baselinestretch}{1.0}
\small\normalsize

\scriptsize

\bibliographystyle{IEEEtran}
\bibliography{bib-quantum}

\end{document}